# Exciton Confinement in Two-Dimensional, In-Plane, Quantum Heterostructures

*Gwangwoo Kim[1], Benjamin Huet[2], Christopher E. Stevens[3,4], Kiyoung Jo[1], Jeng-Yuan Tsai[5], Saiphaneendra Bachu[6], Meghan Leger[6], Kyung Yeol Ma[7], Nicholas R, Glavin[8], Hyeon Suk Shin[7], Nasim Alem[2,6], Qimin Yan[5], Joshua R. Hedrickson[3], Joan M. Redwing[2,6], Deep Jariwala[1,\*]*

[1]*Department of Electrical and Systems Engineering, University of Pennsylvania, Philadelphia, PA 19104, USA.*

[2]*2D Crystal Consortium-Materials Innovation Platform, Materials Research Institute, The Pennsylvania State University, University Park, PA 16802, USA.*

[3]*Air Force Research Laboratory, Sensors Directorate, Wright-Patterson Air Force Base, OH 45433, USA.*

[4]*KBR Inc., Beavercreek, OH 45431, USA.*

[5]*Department of Physics, Northeastern University, Boston, Massachusetts 02115, USA.*

[6]*Department of Materials Science and Engineering, The Pennsylvania State University, University Park, PA 16802 USA*

[7]*Department of Chemistry, Ulsan National Institute of Science and Technology (UNIST), UNIST-gil 50, Ulsan 44919, Republic of Korea.*

[8]*Air Force Research Laboratory, Materials and Manufacturing Directorate, Wright-Patterson Air Force Base, OH 45433, USA.*

**\*Corresponding Authors**

*E-mail Addresses:* dmj@seas.upenn.edu

**Abstract**

Two-dimensional (2D) semiconductors are promising candidates for optoelectronic application and quantum information processes due to their inherent out-of-plane 2D confinement. In addition, they offer the possibility of achieving low-dimensional in-plane exciton confinement, similar to zero-dimensional quantum dots, with intriguing optical and electronic properties via strain or composition engineering. However, realizing such laterally confined 2D monolayers and systematically controlling size-dependent optical properties remain significant challenges. Here, we report the observation of lateral confinement of excitons in epitaxially grown in-plane $MoSe_2$ quantum dots (~15-60 nm wide) inside a continuous matrix of $WSe_2$ monolayer film via a sequential epitaxial growth process. Various optical spectroscopy techniques reveal the size-dependent exciton confinement in the $MoSe_2$ monolayer quantum dots with exciton blue shift (12-40 meV) at a low temperature as compared to continuous monolayer $MoSe_2$. Finally, single-photon emission was also observed from the smallest dots at 1.6 K. Our study opens the door to compositionally engineered, tunable, in-plane quantum light sources in 2D semiconductors.

**Introduction**

Exciton confinement in low-dimensional materials modifies the density of states and enhances the Coulomb interaction between electrons and holes, resulting in a range of novel effects for both fundamental physics and device applications. Over the past decade, atomically thin two-dimensional (2D) crystals have been extensively explored in developing quantum optical devices. However, the lack of lateral confinement of excitonic wave functions in such structures has limited their potential for quantum applications. Single photon emission from point defects[1-6] and localized strains[7-11] in the 2D crystals have been extensively reported with optically detected magnetic resonance[12-14], such as spin coherence, spin relaxation time, and the nature of spin-spin interactions. However, deterministic positioning of individual defects and achieving uniform emission wavelength have remained a frontier challenge in this field. Several groups have also attempted to solve this challenge of individual exciton confinement in transition metal dichalcogenides (TMDs) monolayers by physically shaping them into quantum dots (QDs) via both top-down[15-19] and bottom-up processes[20-22]. Many such attempts have demonstrated that decreasing the lateral dimensions of the TMD QDs results in a distinct blue shift in both emission and absorption spectra, widely recognized as a signature of quantum confinement[15, 19-22]. However, in all cases, these QDs are produced in a manner such that their one-dimensional (1D) edges are exposed. This leads to edge oxidation or covalent chemistry with other functional groups resulting in energy levels and trap states that affect radiative recombination rates and cause a broader distribution of electronic states. Therefore, achieving a seamless and defect-free interface between 2D QDs and matrix materials within an in-plane 2D combination is a significant milestone that remains to be achieved. Further, even though some evidence of lateral quantum confinement has been observed before, demonstration of single-photon quantum emission from compositionally confined 2D QDs remains unachieved.

In this study, we have successfully demonstrated the lateral confinement of excitons in large area 2D MoSe$_2$ QD@WSe$_2$ matrix heterostructures grown by a metal organic chemical vapor deposition (MOCVD) method. These heterostructures were created using sequential epitaxial growth to achieve an ultraclean interface. By controlling the reaction time, we can manipulate the size of the triangular MoSe$_2$ QDs in the range of 15-60 nm. Our optical spectroscopic measurements prove size-dependent exciton confinement within the MoSe$_2$ monolayer QDs. Further, our confined heterostructures exhibited quantum emission with ~0.6 nm spectral line width at cryogenic temperatures for dots as small as 10 nm and a single photon purity of $g^2(0)$ = ~0.4. Our results serve as an important milestone in achieving quantum-confinement and quantum emission in bottom-up grown 2D QDs at scale opening new avenues for exploring confined excitonic physics and developing novel quantum photonic devices.

**Results**

**Fabrication and characterization of 2D quantum heterostructures**

**Figure 1a** schematically illustrates the fabrication process for the single-layer $MoSe_2$ quantum dots (QDs) embedded in a single-layer $WSe_2$ matrix by a sequential epitaxial growth using a horizontal MOCVD system. As a first step, the triangular $MoSe_2$ QDs are grown on a c-plane sapphire substrate at a growth temperature of 950 °C for short reaction times (1, 5, 10 min) using $Mo(CO)_6$ metal precursor and $H_2Se$ chalcogen gas (see Methods for details of MOCVD conditions). Note that the size of the $MoSe_2$ QDs can be varied by adjusting the growth time, and this aspect will be discussed further in characterization. After the $MoSe_2$ QD growth, we sequentially grow a single-layer $WSe_2$ matrix around the QDs in the same chamber without taking out the samples. The growth of the $WSe_2$ monolayer is carried out at the same temperature while supplying $W(CO)_6$ metal precursor and keeping the $H_2Se$ throughout the growth to minimize decomposition of the QDs. A strict 2D in-plane growth along the edges of the $MoSe_2$ QDs can be attained in this manner. Before cooling down, the in-plane heterostructure is further exposed to the chalcogen source to heal any vacancies that could have happened during the growth and prevent the formation of vacancies via hydrogen etching.

The atomic structure of the $MoSe_2$ QDs embedded in the $WSe_2$ matrix was examined using transmission electron microscopy (TEM). **Figure 1b** shows a low-magnification annual dark-field scanning TEM (ADF-STEM) image of the in-plane $MoSe_2$ QDs@$WSe_2$ heterostructures after the film was detached from the sapphire substrate and transferred to a TEM grid (see Methods for details). The presence of the $MoSe_2$ QDs embedded in the $WSe_2$ matrix is confirmed owing to the $Z$ contrast mechanism of the ADF-STEM imaging technique[23], wherein the $MoSe_2$ QDs appear darker compared to the $WSe_2$ matrix. It is worth noting that the size of the MoSe$_2$ QDs can be reduced to 15-60 nm by controlling growth time (**Supplementary Figure S1**). Moreover, an atomic-resolution ADF-STEM image (**Figure 1c**) obtained from the interface illustrates that the atomically sharp interface formed between the QD and the matrix ensures the in-plane heteroepitaxy of the 2D heterostructures. This is further supported by the clear spatial separation of the Mo and W signals in the energy dispersive spectroscopy (STEM-EDS) maps (**Figure 1d-g**).

In addition, the formation of the MoSe$_2$ QDs was confirmed using micro-Raman spectroscopy with a 633 nm laser on the heterostructures transferred onto SiO$_2$ substrates (**Supplementary Figure S2**). The Raman spectra of the in-plane quantum heterostructures exhibited typical Raman signals of MoSe$_2$ and WSe$_2$, including the characteristic A$_{1g}$ symmetry Raman modes of MoSe$_2$ (red) and WSe$_2$ (blue) at 242 and 250 cm$^{-1}$, as well as WSe$_2$ resonance peak (green) with 633 nm laser at 239 cm$^{-1}$. All Raman modes showed no change in the peak position depending on the QDs growth time, but the intensity of both A$_{1g}$ peaks varied. **Supplementary Figure S2c** depicts the plot of A$_{1g}$ intensity ratio on the MoSe$_2$ QDs@WSe$_2$ matrix as a function of the growth time, which indicates a substantial drop in the intensity ratio with the QDs population. These findings are consistent with the increase in the average size and density of the QDs as the growth time is prolonged.

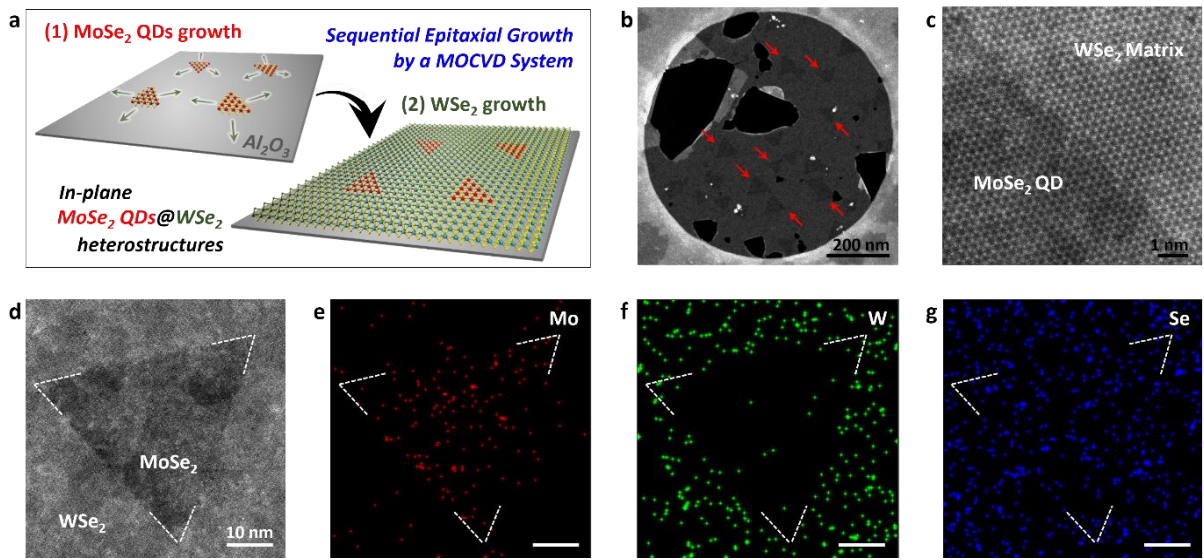

**Fig. 1. Sequential epitaxial growth of the in-plane MoSe₂ QDs@WSe₂.** (**a**) Schematic representation of sequential epitaxial growth of in-plane QDs heterostructures via MOCVD. (**b**) Low magnification ADF-STEM image of the heterostructures consisting of the 5 min-MoSe₂ QDs and the WSe₂ matrix. The MoSe₂ QDs are marked with the red arrows. (**c**) Atomic-resolution ADF-STEM image showing an interface between the MoSe₂ QDs and the WSe₂ matrix (bottom left and top right, respectively). (**d**) ADF-STEM image of the heterostructures with the 5 min-MoSe₂ QDs. (**e-g**) Corresponding STEM-EDS element maps of Mo (**e**), W (**f**), and Se atoms (**g**) for the Figure **d**. The boundaries between the MoSe₂ QD and the WSe₂ matrix are marked with a white dot line from Figure **d**.

## Nano-optical and electrical imaging of the quantum heterostructures

The far-field Raman spectroscopic results described above sample an ensemble of QDs in the WSe₂ matrix since the laser spot size (~1 um) is much larger than the average diameters or spacing between the QDs. Therefore, it is difficult to display the actual QD size and distribution via purely optical means because of the resolution limitations of the instrument. To determine the spatial position and study the band alignment between the MoSe₂ QDs and the WSe₂ matrix, we perform tip-enhanced Raman spectroscopy (TERS) combined with Kelvin probe force microscopy (KPFM) on the heterostructures. First, an epoxy-assisted template stripping procedure was used to prepare the 5min-MoSe$_2$ QDs@WSe$_2$ heterostructures buried in a metallic Au substrate (**Figure 2a**, see Methods in details) which enhances optical signals due to the high local electric field and the Purcell effect for Raman scattering phenomena[24, 25]. As shown in the atomic force microscopy (AFM) height image (**Figure 2b**), despite a slight interface gap between the heterostructures and the Au template, they displayed a smooth topography with minimal surface roughness (Heterostructure: 0.09 nm, Au: 0.13 nm). This suggests that the flat sample prepared by the Au stripping process offers more spatially uniform contact. KPFM mapping of the quantum heterostructures in **Figure 2c** exhibits a noticeable contrast in the contact potential difference (CPD) between the MoSe$_2$ QDs (dark, marked with black dotted circles) and the surrounding WSe$_2$ matrix (bright). This potential difference (~40 mV) arises due to the disparity in work functions between the two materials[26]. Although we measured a smaller difference than expected (~190 meV) due to the quantum-confined dots structure, the difference in potential allows the two materials to be clearly distinguished. The built-in potential of the quantum heterostructure will drive the electrons to diffuse from the WSe$_2$ matrix to the MoSe$_2$ QDs along the direction of the built-in field, demonstrating its characteristic as type II semiconductor heterostructures[27, 28]. Additionally, it is noted that the size of the MoSe$_2$ QDs embedded in the WSe$_2$ matrix was found to be 50 nm (**Figure 2d**), which is consistent with the size observed from the TEM measurement (**Figure 1b**). TERS spectra were collected from the MoSe$_2$ QDs (red) and the WSe$_2$ matrix (blue) regions in **Figure 2f**. Unlike the previous far-field measurement using a 633 nm laser, a 785 nm laser used for gap mode TERS measurement between the tip and the underlying Au template was weakly resonant with both materials. The TERS spectra from the MoSe$_2$ QD and the WSe$_2$ matrix exhibited several features similar to resonant far-field Raman peaks (**Supplementary Figure S2**). Still, we analyzed the out-of-plane A$_{1g}$ mode, which was prominent by signal enhancement in the gap mode. Interestingly, the corresponding TERS maps in the ranges of 300-307 cm$^{-1}$ (WSe$_2$ A$_1$ mode, **Figure 2e**) and 240-244 cm$^{-1}$ (MoSe$_2$ A$_{1g}$ mode, **Figure 2g**) allowed for the identification of the 50 nm-sized MoSe$_2$ QDs embedded in the WSe$_2$ matrix. By using different color channels to render the intensity for the WSe$_2$ (blue) and the MoSe$_2$ QDs (red), a color-integrated image (**Figure 2f**) shows the spatial composition distribution in these in-plane quantum heterostructures.

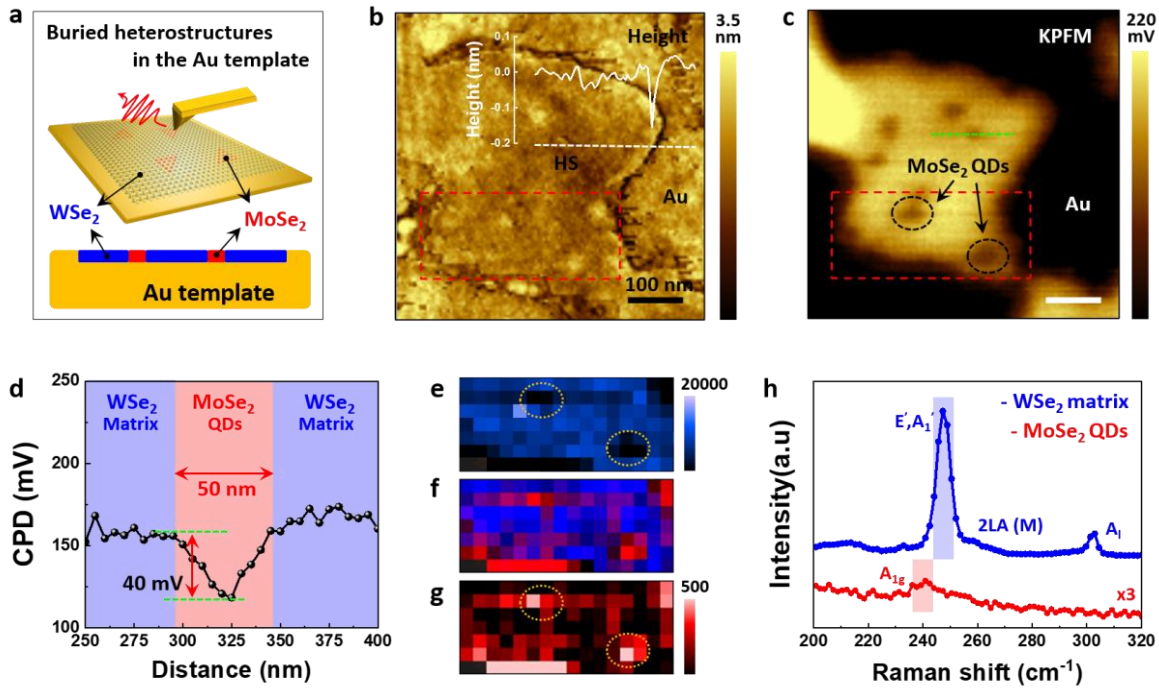

**Figure 2. Nanoscale optical and electrical imaging of the MoSe$_2$ QDs embedded in the WSe$_2$ matrix.** (**a**) Schematic representation of nanoscale scanning probe techniques (top) on the 5min-MoSe$_2$ QDs@WSe$_2$ heterostructures buried in an Au templates (bottom: cross-

sectional view). The samples are prepared by Au-assisted transfer process[24, 25]. (**b, c**) AFM height image and surface potential map of the heterostructures consisting of the 50 nm-sized MoSe$_2$ QDs and the WSe$_2$ matrix. (**Inset**) Height profile marked with white line in Figure **b** showing no discernible height difference across the crystal proving truly in-plane epitaxy (embedding) of the MoSe$_2$ QD in the WSe$_2$ matrix. The MoSe$_2$ QDs are marked with black dotted circles. (**d**) Surface potential profile following the green dotted line in Figure **c**. (**e-g**) TERS spatial maps of the quantum heterostructures following the red dotted squares in Figures **b** and **c**. The TERS images were created within the spectral ranges of 245-255 cm$^{-1}$ (**e**, WSe$_2$ A$_1$ mode) and 235-245 cm$^{-1}$ (**g**, MoSe$_2$ A$_{1g}$ mode) with a step size of 20 nm. The MoSe$_2$ QDs are marked with yellow dotted circles. (**f**) Overlaid image of Figures (**e**) and (**g**). (**h**) TERS spectra of the MoSe$_2$ QD regions (red) and the surrounding WSe$_2$ matrix (blue) as highlighted in the TERS map.

**Exciton confinement in the quantum heterostructures**

Thin hexagonal boron nitride (h-BN) films (**Supplementary Figure S3**) were employed to encapsulate the top and bottom of the quantum heterostructures, as shown in **Figure 3a**, creating stable structures with confined excitons and protecting the samples from unwanted contamination during the measurements. The large area-h-BN tri-layers were grown on a c-plane sapphire substrate by a CVD method[29]. These tri-layers were then stacked vertically with the in-plane heterostructures, using a polymethyl methacrylate (PMMA)-assisted wet-transfer technique on a SiO$_2$ substrate (see the detailed process in **Supplementary Figure S4**). Note that, in the transfer process, we modified the order of a layer-by-layer stacking to minimize the PMMA residue between the layers. The PMMA-coated top h-BN layers were first transferred to the as-grown quantum heterostructures on the sapphire substrate, followed by coating the samples with PMMA again and then transferring them onto the bottom h-BN film. This method allowed residues to be left on the h-BN while ensuring that clean interfaces were established without residues on either side of the quantum heterostructures. Finally, the stacked samples were transferred onto the SiO$_2$ substrate and annealed at 300 °C with Ar flow (50 sccm) in a vacuum tube furnace. This facilitated better contact by eliminating any trapped solvent or gas molecules at the interface between each layer.

The confinement of excitons in the encapsulated quantum heterostructures was examined using micro-photoluminescence (PL) spectroscopy with a 633 nm-excitation laser. **Figure 3b,c** show the temperature-dependent PL spectra of the MoSe$_2$ QDs@WSe$_2$ quantum heterostructures encapsulated in h-BN films, where the MoSe$_2$ QDs were grown for 5 min (**Figure 3b**) and 10 min (**Figure 3c**). The PL peaks showed a considerable shift to higher energies with decreasing temperature due to the lattice shrinkage and reduced electron-phonon interaction at low temperatures[30]. At room temperature (red spectra in **Figure 3b** and **c**), the PL intensities of the MoSe$_2$ QDs on both samples were not large due to the small size and low density of the QDs. However, at ~80K, the emission signal from the QDs signal is clear as compared to that of the WSe$_2$ matrix. The PL intensities (neutral exciton + charged exciton, $X^0+X^+$) ratio of MoSe$_2$ to WSe$_2$ is plotted in **Supplementary Figure S5b**, showing strong enhancement at low temperature, possibly due to dominant electron transfer from the WSe$_2$ matrix to the MoSe$_2$ QDs by the band structure modulation[31]. Furthermore, the confined excitons in the heterostructures can be modulated through electrostatic gating. We fabricated a gate-tunable device using mechanically exfoliated graphene and h-BN flakes as the top-gate electrode and the dielectric, respectively. By adjusting the Fermi level in two materials with Type II band alignment, we further demonstrate via emission spectroscopy that efficient electron transfer can be controlled via gating in these in-plane quantum heterostructures (see details in **Supplementary Figure S6**).

The quantum confinement of excitons in the MoSe$_2$ QDs can also be confirmed by comparing their PL energy positions with those of large-area MoSe$_2$ and WSe$_2$ monolayers grown on sapphire substrates in the same MOCVD system. The PL energy position of the neutral exciton of MoSe$_2$ (right) and WSe$_2$ (left) on the quantum heterostructures (MoSe$_2$ QD growth time of 5 min: red, 10 min: blue) and reference monolayers (MoSe$_2$ and WSe$_2$: black) were compared as shown in **Figure 3d**. Depending on the samples, there was no change in the WSe$_2$ PL position. Still, the excitonic features of the MoSe$_2$ PL were blue-shifted on the QD samples, indicating the exciton confinement. The confinement effect was larger in the 5 min-MoSe$_2$ QDs heterostructures (~20 meV) than in the 10 min sample (~15 meV), indicating a stronger blue shift with a smaller lateral size of the QDs. It is worth noting that although the QDs lateral size is larger than the exciton Bohr radius (~1.5 nm) in the TMD monolayers[32, 33], the excitons can still be in a weak lateral confinement regime[15, 19-22]. Theoretical calculations will be discussed below to explore this confinement effect further.

We performed first-principle calculations based on density functional theory (DFT) to investigate the shift of optical transitions due to the confinement effects in QDs with various sizes. Equilateral triangular MoSe$_2$ quantum dots embedded in WSe$_2$ with edge lengths ranging from 2.7 to 5.3 nm were studied. As the size of the QDs decreases, the absolute energy of the conduction band minimum (CBM) rises due to the quantum confinement effect, while that of the valence band maximum (VBM) decreases at a much slower rate. The energy difference between the CBM and the VBM increases with increasing dot size (**Supplementary Figure S7**). For instance, in the 2.7-nm quantum dot, the projected density of states reveals that the CBM is mainly contributed by the quantum dot, with the corresponding wave function being spatially localized within the dot (**Supplementary Figure S8**). Note that the electronic state at the VBM is delocalized, while the state at 0.25 eV below the VBM turns out to be confined by the quantum dot. This observation is consistent with the type-II band alignment between $MoSe_2$ and $WSe_2$, which effectively forms quantum dots for electrons but does not imply confinement for those holes at the VBM.

Next, we evaluate the effect of quantum confinement on the intra-dot optical transition by defining an energy shift between two relevant energies: (1) the energy difference between the CBM of QDs and the VBM of pristine $MoSe_2$; (2) the optical gap of pristine $MoSe_2$. To account for the excitonic effects and the underestimation of band gaps by DFT, we increase the calculated band energy differences by 0.11 eV (which is the energy difference between the calculated band gap and the measured optical gap of $MoSe_2$) to roughly evaluate the optical transition energies. The energy shift is plotted as a function of the inverse size of QDs. Due to the limitations in computational capacity, linear extrapolation is used to estimate the results for more sizable QDs. The energy shift is estimated to be 40, 20, and 12 meV for 15, 30, and 50-nm quantum dots, respectively (**Figure 3e**). The predicted shift of transition energies in 30- and 50-nm quantum dots are comparable with the observed energy shifts of 20 and 15 meV in the 5 min- and 10 min-$MoSe_2$ QDs (**Figure 3d**). It is noted that the discussion of results pertaining to the 1 min-$MoSe_2$ QDs will be addressed in more detail later. Although our calculations do not explicitly include the excitonic effects, which can be significant in two-dimensional materials, the consistent results suggest that the change in the band energies of those electronic states involved in optical transitions may be dominant in the observed shift of transition energies. This finding is consistent with the conclusions of previously reported theoretical[19, 34] and experimental results[15, 19, 35-37] in the 2D TMD QDs.

In addition to PL analysis, the confinement effect in the quantum heterostructures was also verified through reflectance measurements (**Figure 3f** and **Supplementary Figure S9**). **Figure 3f** displays the reflectance spectra of the 10 min-MoSe$_2$ QDs heterostructures (black) and reference monolayers (MoSe$_2$: red, WSe$_2$: blue) obtained at ~80 K. As observed in the PL results, the confinement of the MoSe$_2$ QDs is evident in the reflectance spectra, while the positions of WSe$_2$ absorption peaks are nearly identical between the heterostructures and the WSe$_2$ monolayer. These two optical measurements provide clear proof of the optical confinement effect.

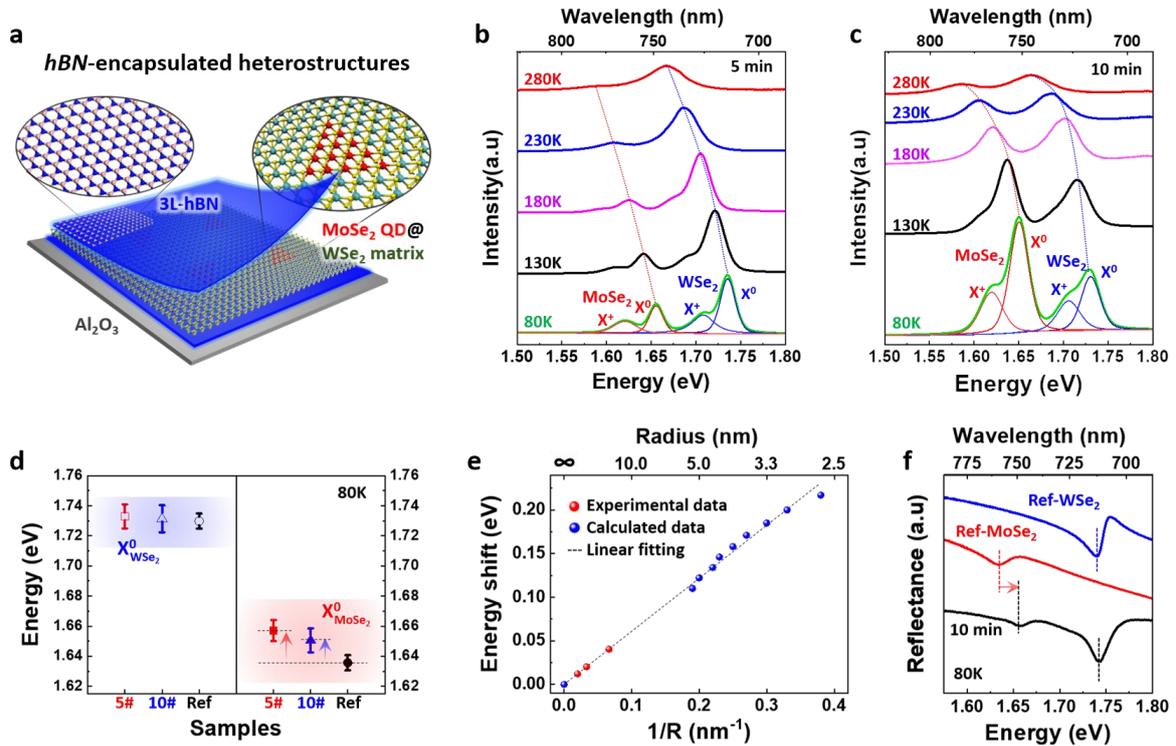

**Figure 3. Exciton confinement of the MoSe$_2$ QDs embedded in the WSe$_2$ matrix.** (**a**) Schematic representation of the in-plane MoSe$_2$ QDs@WSe$_2$ heterostructures encapsulated in top and bottom h-BN tri-layers. The Mo, W, Se, B, and N atoms are represented in red, green, yellow, pink, and blue, respectively. (**b**, **c**) Temperature-dependent PL spectra (280 K to 80 K)

of the heterostructures with different growth times of the MoSe$_2$ QDs (**b**: 5min, **c**: 10 min). These spectra were obtained by continuous 633 nm-laser with an excitation power of 20 µW and a 50x lens with 0.35 NA. (**d**) Comparison of PL energy position of the main neutral excitons of MoSe$_2$ and WSe$_2$ in the heterostructure samples (red: 5 min, blue: 10 min) with reference monolayers (black). The points are plotted from the peak positions measured at 80 K in Figure **b** and **c**. The reference MoSe$_2$ and WSe$_2$ monolayers are prepared in the same MOCVD chamber. (**e**) Relation between the energy shifts on size of the MoSe$_2$ QDs embedded in the WSe$_2$ matrix. The dashed blue line represents the linear fit of the energy shifts with respect to the inverse size of QDs. The blue dots represent the computationally estimated energy shifts in the QDs with sizes ranging from 2.7 to 5.3 nm. The three red dots are predicted energy shifts in 15, 30, and 50-nm quantum dots using linear extrapolation. The experimentally determined optical band gap of pristine MoSe$_2$ is set as zero of energy and associated with an infinitely large (R = ∞) quantum dot. (**f**) Reflectance spectra of the heterostructures (10 min-MoSe$_2$ QD: black) and reference monolayers (MoSe$_2$: red, WSe$_2$: blue) measured at 80 K.

In our previous analysis of temperature-dependent PL, a stronger confinement effect was expected on the heterostructures of the 1 min growth time-MoSe$_2$ QDs with smaller sizes (approximately 15 nm). However, due to the limited size and density of the quantum dots, it was challenging to detect the MoSe$_2$ PL signal (**Supplementary Figure S10**). This was mainly because the confocal Raman system, with a nitrogen-cooling stage with a large beam size from a 50x, 0.35 NA lens, was not able to probe and detect the small dots effectively. Therefore, to address this limitation and directly explore the quantum confinement on the QD heterostructures, we performed cryogenic PL measurements (down to 1.6 K) with a 100x lens with 0.82 NA. **Figure 4a** lists the temperature-dependent PL characteristics of the 1 min-MoSe$_2$ QDs sample, showing a clear MoSe$_2$ PL signal. A gradual blue-shift of the peaks (MoSe$_2$ QD, WSe$_2$, and defect states) with decreasing temperature is clearly observed, as in the previously shown **Figure 3b-c**.

Additionally, we had also observed a distinct and sharp MoSe$_2$ PL emission when the temperature approached 1.6K. We hypothesize that they originate from the quantum-confined main excitons of the MoSe$_2$ QDs because of the matching energy values as discussed above. **Figure 4b** shows the PL spectra on the quantum heterostructures (1 min growth time-MoSe$_2$), which were obtained with a pulsed excitation laser (640 nm, 1 MHz) with 100 nW power at 1.6 K showing an emission with a narrow line width (612 ± 124 μeV) at an energy of 1.675-1.692 eV. However, this narrow emission also resembles single photon emitters from dark excitons and defect states in WSe$_2$ monolayers, as reported in several prior studies[1, 2, 7, 11]. To eliminate the possibility for these narrow emission lines to be attributed to defect states of the WSe$_2$ matrix, we prepared and measured MOCVD-grown pure 2D WSe$_2$ and MoSe$_2$ monolayers in the same chamber with identical growth conditions. As shown in **Supplementary Figure S11b**, the emission from the defect state of WSe$_2$ was observed at a slightly lower energy position (1.65-1.66 eV), and it has a quite broad band, as compared to the quantum heterostructures. Further, the main neutral exciton in pure MoSe$_2$ samples was observed at 1.652 eV (**Supplementary Figure S11a**, bottom). It is worth noting that the quantum emission observed on the heterostructures (**Figure 4d**, inset) was blue-shifted by ~40 meV due to the confined excitons in the quantum dots, which is consistent with the theoretically predicted value in **Figure 3e**.

To further validate that the emission acts as a truly quantum light source, we measure the second-order correlation function g$^2$(t) using a Hanbury-Brown-Twiss (HBT) set-up with two single photon-counting avalanche photodiodes (APDs). **Figure 4c** shows the second-order correlation under continuous-wave (CW) excitation of the emitter on one of the selected sharp emission lines binned in between the dotted lines (**Figure 4d**, inset) using a broad band-pass filter and tunable short and long pass filters. By fitting the measured data with a standard two level antibunching function, we calculate a $g^2(0) = 0.4 \pm 0.02$, which drops below the threshold for a single quantum emitter of 0.5 (more spectra on several other positions on the sample are provided in **Supplementary Figure S12**). This confirms that the $MoSe_2$ QD is truly a single photon emitter. **Figure 4d** shows the PL lifetime of the $MoSe_2$ QDs on the heterostructures, which was measured by excitation with a 640 nm, ~200 ps, 10MHz pulsed diode laser and sending the spectrally filtered output around the quantum dot wavelength to a single-photon-counting APD. The measured long lifetime of ~ 3 ns from the exponential fitting (black line) is consistent with the behavior shown by typical III-V semiconductor quantum dots[38-41].

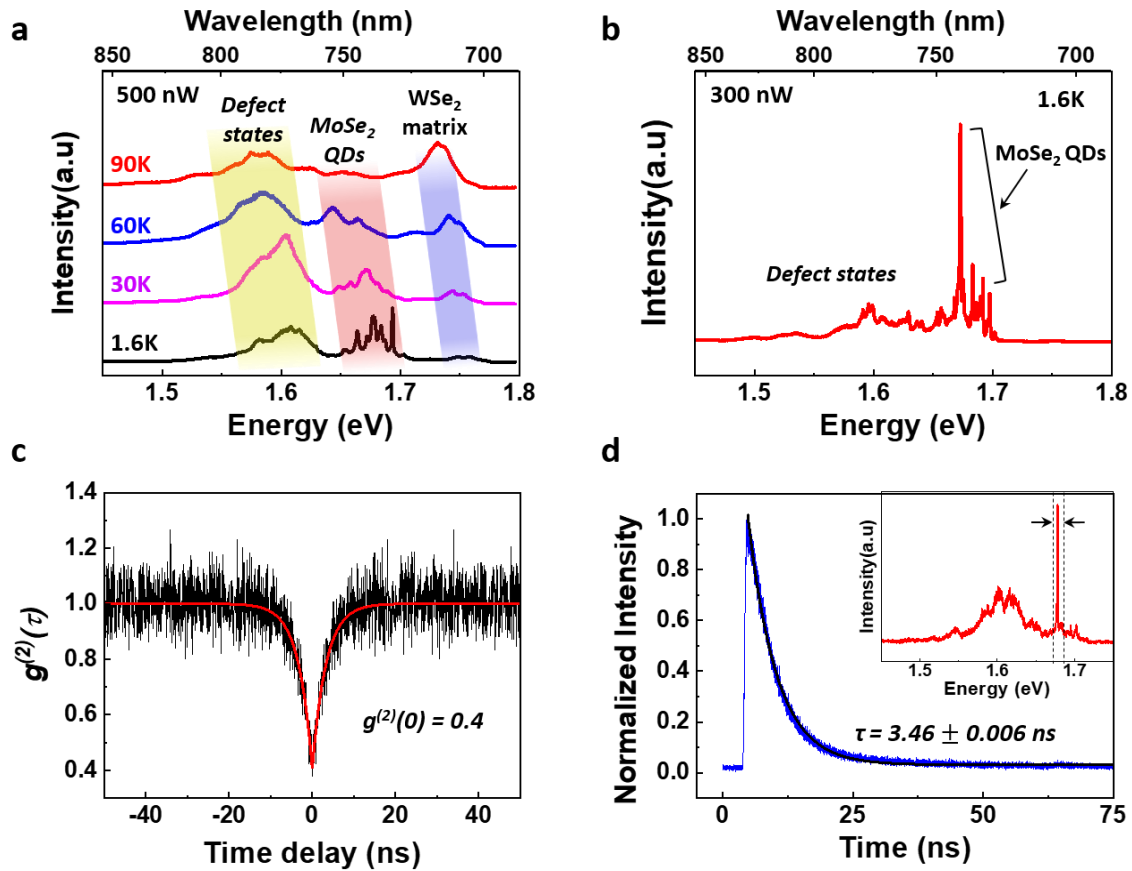

**Figure 4. Cryogenic PL measurement and single photon emission in the quantum heterostructures.** (**a**) Temperature-dependent PL spectra (90 K to 1.6 K) of the heterostructures with the 1 min growth time-$MoSe_2$ QDs, measured using a pulsed laser excitation (500 nW, 640 nm, 1 MHz). The emission peaks from defect states, $MoSe_2$ QDs, and $WSe_2$ matrix are marked with yellow, red, and blue regions, respectively. (**b**) Representative cryogenic PL spectrum of the QD heterostructures with an excitation power of 100 nW at 1.6 K. (**c**) Second-order photon correlation curve for the PL signal of the $MoSe_2$ QDs. (**d**) TRPL spectra for the sharp emission of the $MoSe_2$ QDs (1.688 eV), shown in Figure **b**. The time-resolved PL data (blue line) are convoluted (black line) with the instrument response function, using an exponential function $I = A * exp(-t/\tau)$.

**Conclusions**

In conclusion, we have successfully demonstrated the lateral confinement of excitons via compositionally controlled in-plane 2D quantum heterostructures of the $MoSe_2$ QDs embedded in the $WSe_2$ matrix. These heterostructures show quantum-confined emission that is significantly blue-shifted from the main neutral excitons of pure 2D monolayer $MoSe_2$. Further, the wavelength and intensity of the emission can be modulated passively according to the QD size and actively via electrostatic gating since the QDs are embedded in the $WSe_2$ matrix. Our work represents a significant step towards the synthetic control of truly two-dimensional in-plane epitaxial QDs, making them a versatile and tunable quantum light source. To fully exploit their potential for future, high-fidelity quantum light sources, further work must focus on their controllability in terms of spatial position, density, and composition.

**Methods**

**Growth of the in-plane MoSe$_2$ QDs@WSe$_2$ heterostructures**

The in-plane MoSe$_2$ QDs@WSe$_2$ heterostructures were synthesized on sapphire substrates in a horizontal MOCVD reactor[42]. The process involved introducing Mo(CO)$_6$ or W(CO)$_6$ metal precursors and H$_2$Se chalcogen source at a growth temperature of 950 °C. After the heating ramp, a 10 min high-temperature annealing step is carried out under a pure H$_2$ atmosphere to remove surface impurities and stabilize the sapphire surface. The growth of the MoSe$_2$ QDs was initiated by introducing Mo(CO)$_6$ metal precursor (6.1 x 10$^{-3}$ sccm) and H$_2$Se chalcogen source (200 sccm) simultaneously. Pure H$_2$ was used as a carrier gas to transport the metal carbonyl in the chambers via a bubbler which was maintained at a controlled pressure of 760 Torr and temperature of 20 °C. The desired size of the MoSe$_2$ QDs (10-50 nm) is typically adjusted by the growth time (1-10 min). Sequentially, stopping the Mo precursor supply, W(CO)$_6$ precursor (8.7 x 10$^{-4}$ sccm) was introduced to grow the WSe$_2$ matrix around the MoSe$_2$ QDs. During the growth of the heterostructures, the identical flow of H$_2$Se gas continued to be supplied, and the system pressure was maintained at 200 Torr. Finally, after the growth, the furnace was cooled to 300 °C in a mix of H$_2$/H$_2$Se and was then cooled further to room temperature in N$_2$.

**Preparation of buried heterostructure in the Au templates**

A thin gold film (100 nm) was deposited onto the surface of The in-plane MoSe$_2$ QDs@WSe$_2$ heterostructures grown on the sapphire substrate by using an e-beam evaporator (Kurt J. Lesker PVD-75) under a high vacuum. Then, a Si wafer was attached to the outer gold surface using an epoxy resin. Once the epoxy was cured at 80 °C for 2h, the gold-sapphire interface was separated by peeling. The in-plane heterostructures are more strongly bound to the Au film and are thus separated from the sapphire substrate. The process results in the transfer of the heterostructures from the sapphire surface to being inlaid in the gold film, exposing the pristine surfaces of the heterostructures that were previously in contact with the sapphire substrate.

**Device fabrication for electrostatic gating.**

For electrostatic gating device preparation, the mechanically exfoliated h-BN and graphene layers were transferred over the sample by using a polydimethylsiloxane (PDMS)-based dry transfer process to use them as each a dielectric and a top gate electrode. Next, the fabrication of Ti (10 nm)/Au (100 nm) electrode contacts was achieved by using electron beam lithography (Elionix ELS-7500EX) and the e-beam evaporator (Kurt J. Lesker PVD-75). Finally, the samples were cleaned in acetone for the lift-off process.

**Optical and structural characterization.**

Far-field Raman and PL spectroscopy were performed in a Horiba LabRam HR Evolution confocal microscope with 633 nm excitation lasers. The signals were collected through a 50× microscope objective (Olympus SLMPLN, NA = 0.35) for low-temperature measurements (from room temperature to 80 K). Also, for the low-temperature analysis, samples were placed in a Linkam stage with a liquid nitrogen supply while cooling and heating and pumped to $5 \times 10^{-3}$ Torr during the measurement. Additionally, for electrostatic gating, the electrical bias was applied using a Keithley 2450 sourcemeter. An OmegaScope Smart SPM (AIST-NT) setup was used for topography scans. For near-field PL measurements, Au coated OMNI-TERS probes (APP Nano) were used in the identical AFM setup coupled to a far-field Horiba confocal microscope with a 633 nm excitation laser.

For the cryogenic PL (from 80K to 1.6 K), time-resolved PL, and $g^{(2)}(\tau)$ measurements, the sample was placed in a cryostat with an in situ 0.82 NA 100× objective. The excitation spot size was approximately 1 μm. For time-resolved PL, the sample was illuminated with 640 nm, 200 ps light generated from a PicoQuant diode laser. For PL saturation and $g^{(2)}(\tau)$ measurements, the sample was illuminated with 640 nm CW diode laser. The signal was collected in a reflection geometry and routed to a spectrometer for PL measurements. For time-resolved PL and $g^{(2)}(\tau)$, the SPEs were first identified using PL and then spectrally filtered with angle tunable Semrock filters before being sent to two fiber-coupled Si avalanche photodiode (APD) detectors. PL saturation measurements confirmed that all time-resolved PL and $g^{(2)}(\tau)$ measurements of SPEs were taken well below saturation.

For the TEM analysis, the as-grown $MoSe_2$ QDs@$WSe_2$ heterostructure films were transferred to Quantifoil Cu TEM grids using a PMMA-assisted wet transfer method[43]. ADF-STEM imaging and STEM-EDS mapping were performed using a dual-corrected Thermo Fisher Titan$^3$ G2 microscope operated at 80 kV. A semi convergence angle of 30 mrad and a screen current of ~ 50-60 pA were used during the imaging.

**Computational methods**

First-principles calculations are performed by using the VASP code[44] with the $r^2SCAN$[45] metaGGA functional and a plane-wave implementation. Parallel GPU computations are used to accelerate the calculations of these large-scale 2D materials systems. The $MoSe_2$ quantum dots of various sizes are embedded in an 18x18 supercell of $WSe_2$, and calculations are performed using the $\Gamma$ point in the Brillouin zone. The calculated band gaps for pristine $MoSe_2$ and $WSe_2$ are 1.53 and 1.63 eV, respectively. A type-II heterostructure of $MoSe_2$@$WSe_2$ is formed. The CBM and the VBM of $WSe_2$ are higher than those in $MoSe_2$ by 0.35 and 0.25 eV, respectively. A vacuum thickness of 15 Å is set to prevent the interactions between periodic images. The plane-wave cutoff energy is set to 315 eV. The structural relaxations are performed for all systems until the force acting on each ion is less than or equal to 0.02 eV/Å. The convergence criteria for total energies in structural relaxations and self-consistent calculations are $10^{-4}$ eV and $10^{-5}$ eV, respectively.

**Acknowledgements**

D.J. and G.K. acknowledge primary support for this work by the Air Force Office of Scientific Research (AFOSR) FA2386-20-1-4074 and partial support from FA2386-21-1-4063. The MOCVD samples were grown by B.H., M.L. and J.M.R. in the 2D Crystal Consortium Materials Innovation Platform (2DCC-MIP) facility at Penn State which is supported by the National Science Foundation under cooperative agreement DMR-2039351. A portion of the sample fabrication, assembly, and characterization were carried out at the Singh Center for Nanotechnology at the University of Pennsylvania, which is supported by the National Science Foundation (NSF) National Nanotechnology Coordinated Infrastructure Program grant NNCI-1542153. The research performed by C.E.S. at the Air Force Research Laboratory was supported by contract award FA807518D0015. J.R.H. acknowledges support from the Air Force Office of Scientific Research (Program Manager Dr. Gernot Pomrenke) under award number FA9550-20RYCOR059. K.J. was supported by a Vagelos Institute of Energy Science and Technology graduate fellowship. J.-Y.T and Q.Y. acknowledge support from the U.S. Department of Energy, Office of Science, under award number DE-SC0023664. The work of S.B. and N.A. was supported by NSF CAREER DMR-1654107 grant. H.S.S. thanks the support by Basic Science Research Program through the National Research Foundation of Korea (NRF) funded by the Ministry of Education (NRF-2021R1A3B1077184). K.Y.M was supported by the research fund (NRF-2022R1C1C2009666) through the National Research Foundation, Republic of Korea.

**Author contributions**

G.K. and D.J. conceived the measurements and sample fabrication idea/concepts. J.M.R. B.H. and M.L. conceived and performed the MOCVD-growth of the in-plane $MoSe_2$ QDs@ $WSe_2$ heterostructures and reference ($WSe_2$, $MoSe_2$) monolayers. G.K. performed sample preparation and far-field optical characterizations (temperature dependence, gate-tunable PL and reflectance measurement down to 80 K). K.J performed near-field PL and KPFM measurements. Q.Y. and T.J. performed DFT-based first-principles calculations. S.B. and N.A performed scanning transmission electron microscopy characterization. C.E.S. and J.R.H. performed cryogenic PL (down to 1.6 K), time-resolved PL, and $g^{(2)}(\tau)$ measurements. K.Y.M. and H.S.S. provided the h-BN films CVD-grown on the sapphire substrates. G.K. and D.J. wrote the manuscript with inputs from all co-authors.

**Data availability**

All data are available in the paper and Supplementary Information. Growth and standard characterization data associated with the samples used in this study is available via ScholarSphere. This includes substrate preparation and recipe data for samples grown by MOCVD in the 2DCC-MIP facility and standard characterization data including AFM images, room temperature Raman/PL spectra and SEM images on the samples.

# Supplementary Information

## Exciton Confinement in Two-Dimensional, In-Plane, Quantum Heterostructures

Gwangwoo Kim[1], Benjamin Huet[2], Christopher E. Stevens[3,4], Kiyoung Jo[1], Jeng-Yuan Tsai[5], Saiphaneendra Bachu[6], Meghan Leger[6], Kyung Yeol Ma[7], Nicholas R, Glavin[8], Hyeon Suk Shin[7], Nasim Alem[2,6], Qimin Yan[5], Joshua R. Hedrickson[3], Joan M. Redwing[2,6], Deep Jariwala[1,*]

[1] Department of Electrical and Systems Engineering, University of Pennsylvania, Philadelphia, PA 19104, USA.

[2] 2D Crystal Consortium-Materials Innovation Platform, Materials Research Institute, The Pennsylvania State University, University Park, PA 16802, USA.

[3] Air Force Research Laboratory, Sensors Directorate, Wright-Patterson Air Force Base, OH 45433, USA.

[4] KBR Inc., Beavercreek, OH 45431, USA.

[5] Department of Physics, Northeastern University, Boston, Massachusetts 02115, USA.

[6] Department of Materials Science and Engineering, The Pennsylvania State University, University Park, PA 16802 USA

[7] Department of Chemistry, Ulsan National Institute of Science and Technology (UNIST), UNIST-gil 50, Ulsan 44919, Republic of Korea.

[8] Air Force Research Laboratory, Materials and Manufacturing Directorate, Wright-Patterson Air Force Base, OH 45433, USA.

**\*Corresponding Authors**

*E-mail Addresses:* dmj@seas.upenn.edu

**Supplementary Figure S1**

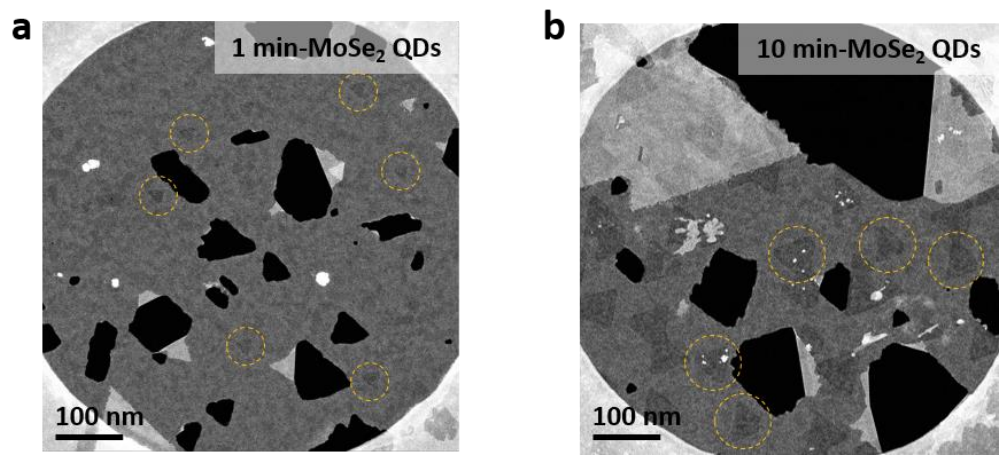

**Figure S1. Comparison of the size of the MoSe$_2$ quantum dots (QDs) in the heterostructures with different growth times of MoSe$_2$ QDs.** (**a**, **b**) Low magnification annual dark-field scanning TEM (ADF-STEM) images of the heterostructures with the MoSe$_2$ QDs (**a**: 1 min, **b**: 10 mins).

**Supplementary Figure S2**

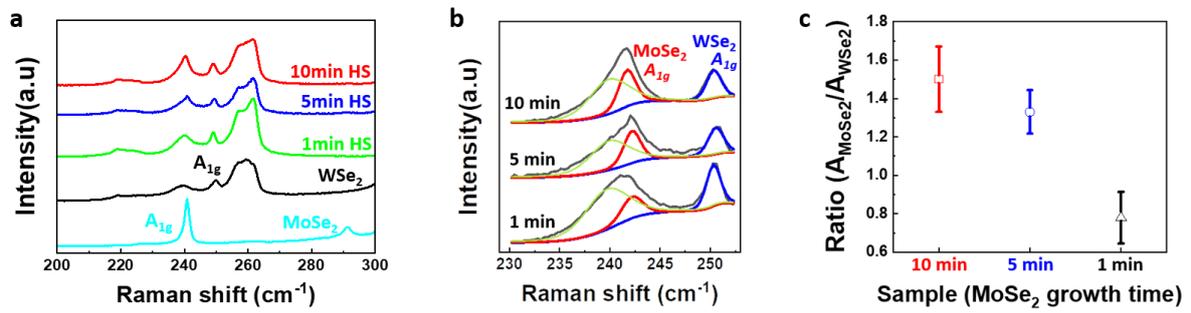

**Figure S2. Raman comparison on the heterostructures with different growth times (1, 5, 10 min) of MoSe$_2$ QDs.** (**a**, **b**) Raman spectra of the heterostructures (10 min: red, 5 min: blue, 1 min: green), WSe$_2$ (black) and MoSe$_2$ monolayers (sky blue). Figure **b** is magnified in the range of 230-255 cm$^{-1}$ in Figure **a**, and each peak is deconvoluted. (**c**) Comparison of Raman area ratio of MoSe$_2$ A1g mode to WSe$_2$ A1g mode on the heterostructures.

**Supplementary Figure S3**

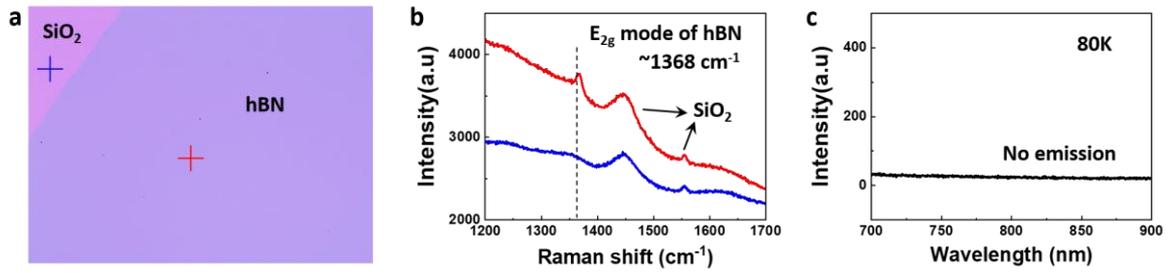

**Figure S3. Characterization of h-BN tri-layers used for encapsulation.** (**a**) Optical microscopic image of the h-BN tri-layers transferred on a SiO$_2$ substrate. (**b**) Raman spectra of the h-BN film (red) and the bare SiO$_2$ substrate (blue) using a 633 nm-laser. The measured points are marked with red and blue crosses in Figure **a**. (**c**) Photoluminescence (PL) spectrum of the h-BN film on SiO$_2$ substrate, which was measured at 80 K.

**Supplementary Figure S4**

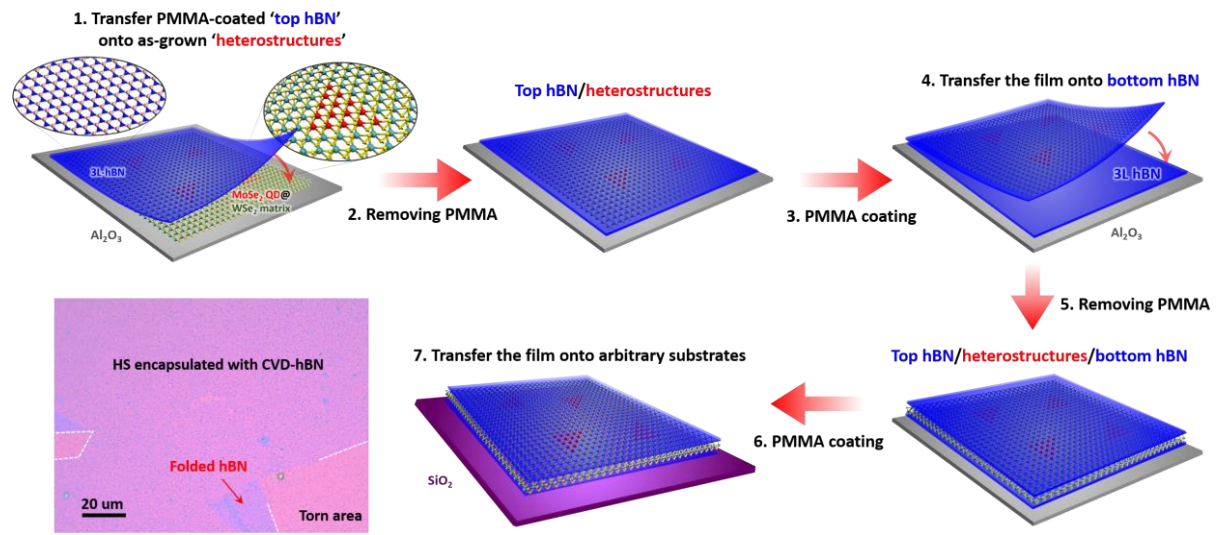

**Figure S4. Fabrication process of the MoSe$_2$ QDs/WSe$_2$ heterostructures encapsulated in top and bottom h-BN layers.** (**Bottom left**) An optical microscopic image of the quantum heterostructures encapsulated with h-BN tri-layers, transferred on the SiO$_2$ substrate.

**Supplementary Figure S5**

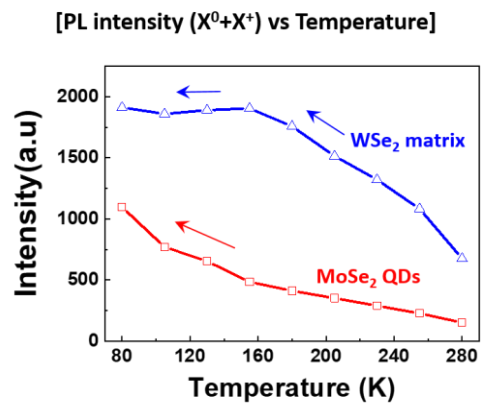

**Figure S5. PL intensity change of MoSe$_2$ QDs (red) and WSe$_2$ (blue) excitons on a 5 min-MoSe$_2$ QDs heterostructures as a function of temperature.** The points were plotted from the spectra in **Figure 3b** and **3c**.

**Supplementary Figure S6**

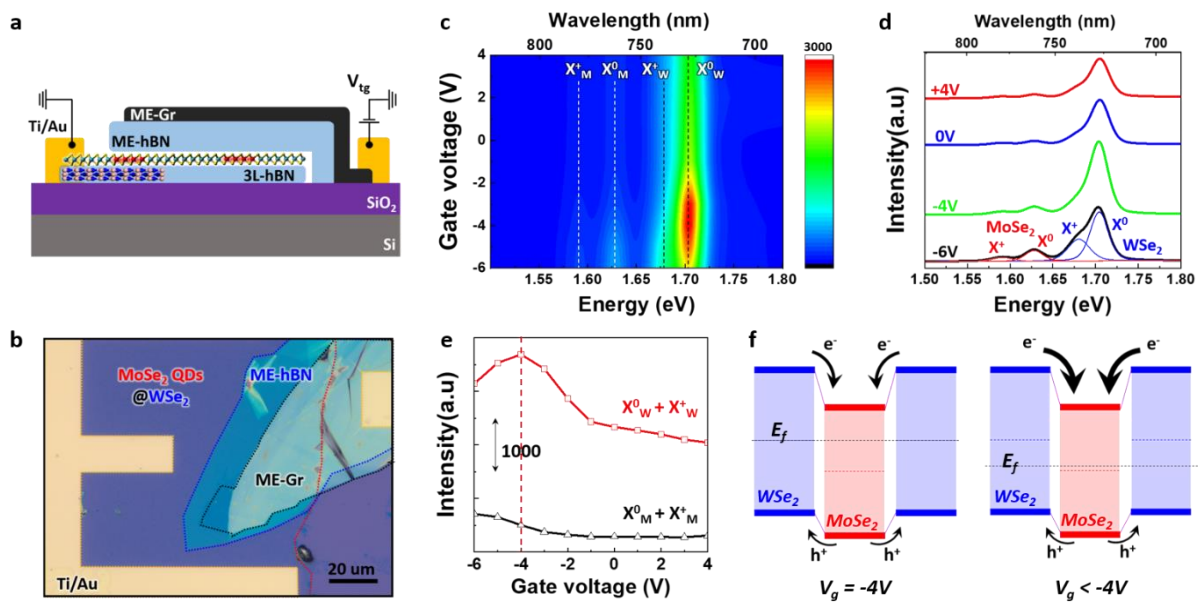

**Figure S6. Gate-dependent PL spectroscopy of the quantum heterostructures.** (**a**, **b**) Schematic (**a**) and optical microscopic image (**b**) of the h-BN-encapsulated quantum heterostructures device with h-BN dielectric layers and graphene top gate electrode. (**c**) PL spectra of the heterostructures at 80 K as a function of the top gate voltage. The color represents the PL intensity. (**d**) PL spectra at the top gate voltage of +4, 0, -4 and -6 V. As the gated voltage sweep decreased from +4V to -6V, there was a corresponding increase in the intensity of $MoSe_2$ and $WSe_2$ trion observed on the heterostructure, indicating the presence of positively charged trion ($X^+$). (**e**) PL intensity change of $WSe_2$ (red) and $MoSe_2$ (blue) PL peaks as a function of gate voltage. Remarkably, the $WSe_2$ signal decreased below -4V gate voltage, indicating that the Fermi level was in the middle of the $WSe_2$ bandgap at a bias of -4 V. However, if a lower bias than this was applied, the signal became smaller due to Pauli blocking as it approached the valence band. (**f**) Schematic band diagram of the quantum heterostructures with the Fermi level modulated by electrostatic gate voltage ($V_g$). The difference in the conduction band edge between the two materials was much larger than that of the valence band. With a decrease in the Fermi level, more electron transfer was induced, leading to further changes in the PL spectra.

**Supplementary Figure S7**

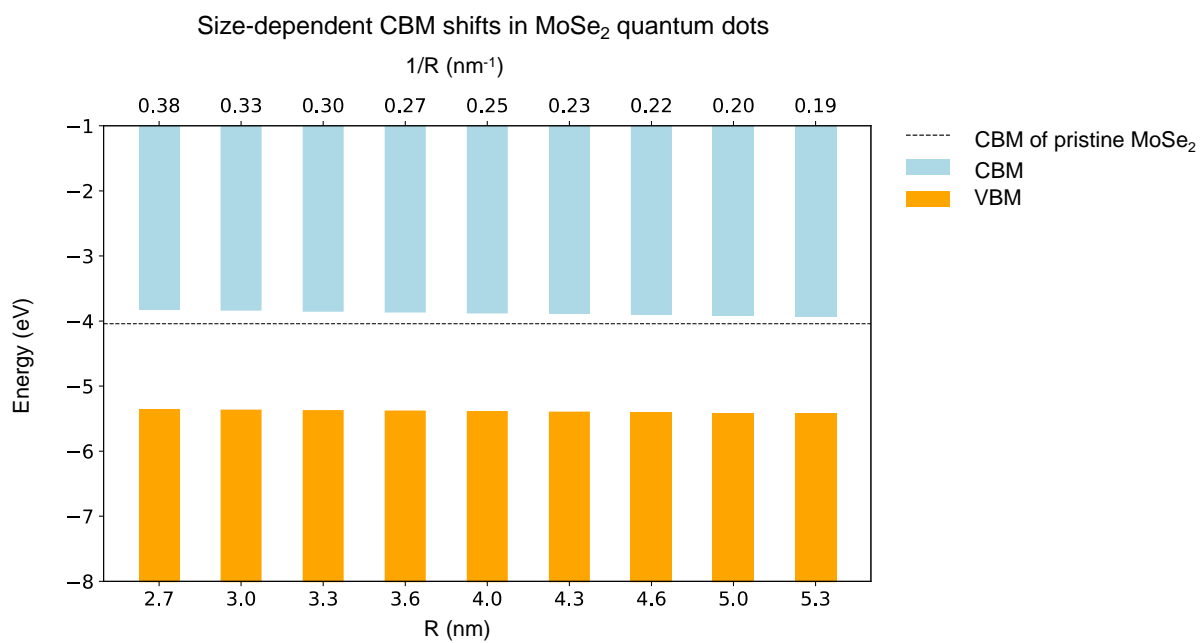

**Figure S7. Size-dependent shift of the conduction band minima (CBM) of MoSe$_2$ QDs embedded in WSe$_2$.** The cyan and orange bars represent the CBM and the valence band maxima (VBM), respectively. The edge lengths of MoSe$_2$ QDs range from 2.7 to 5.3 nm. The dashed line represents the CBM of pristine MoSe$_2$, which is located at -4.04 eV below the vacuum energy.

**Supplementary Figure S8**

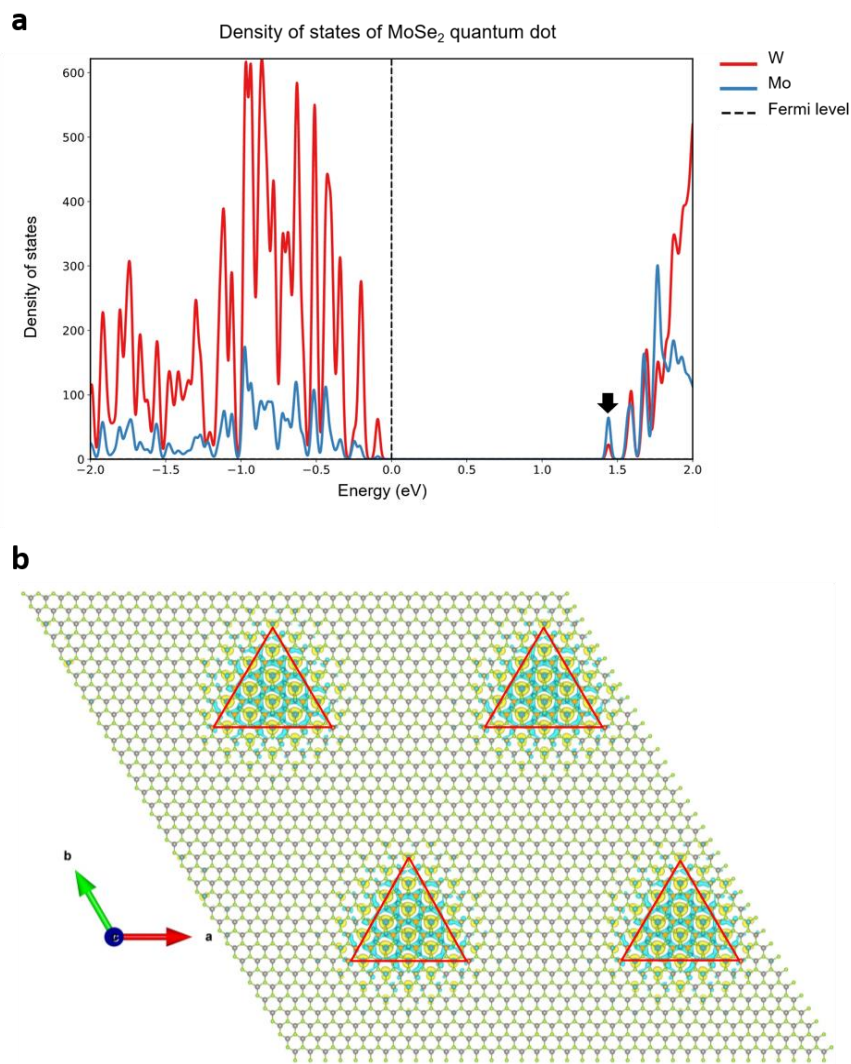

**Figure S8. Projected density of states and the wave function of electronic state at the CBM in a 2.7-nm MoSe$_2$ QDs embedded in an 18x18 supercell of WSe$_2$.** (**a**) The projected density of states showing the contributions of Mo and W (represented by blue and red lines, respectively). The black arrow indicates the CBM. (**b**) The wave function of the electronic state at the CBM exhibits a strong localization within the MoSe$_2$ QDs denoted by the red triangle.

**Supplementary Figure S9**

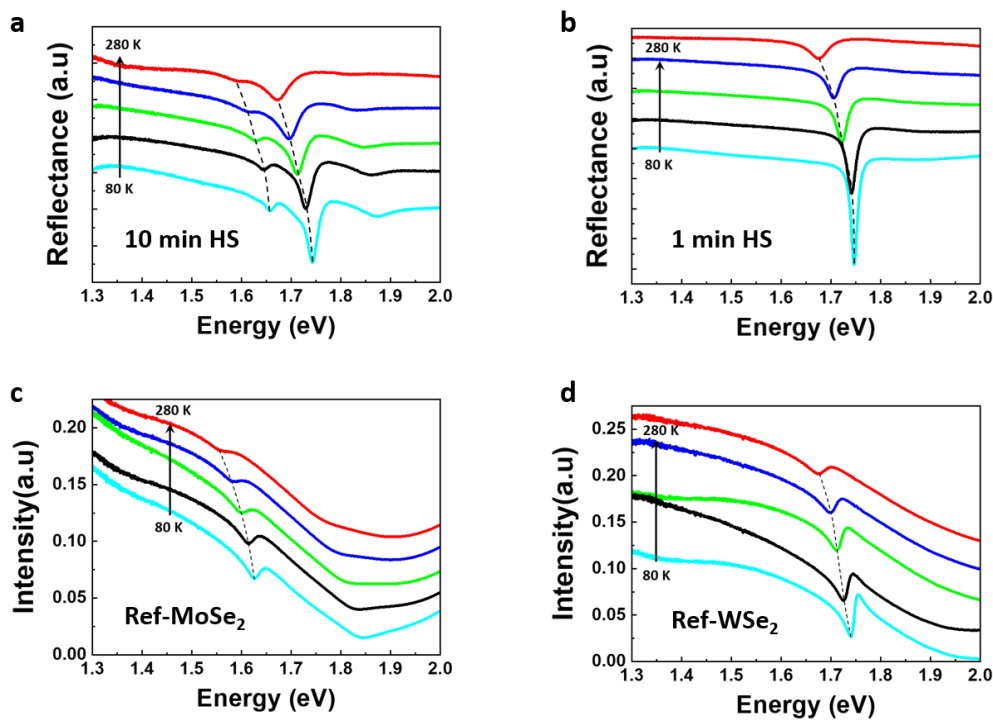

**Figure S9. Temperature-dependent reflectance spectra (280 K to 80 K) of the heterostructures (a: 10 min-MoSe₂ QDs, b: 1 min-MoSe₂ QDs) and reference monolayers (c: MoSe₂, d: WSe₂ monolayers).** The references $MoSe_2$ and $WSe_2$ monolayers are prepared in the same metal organic chemical vapor deposition (MOCVD) chamber. For the 1 min-heterostructure (**b**), the absorption peak of $MoSe_2$ QDs could not be obtained due to the low detection level, similar to PL analysis (**Supplementary Figure S10**).

**Supplementary Figure S10**

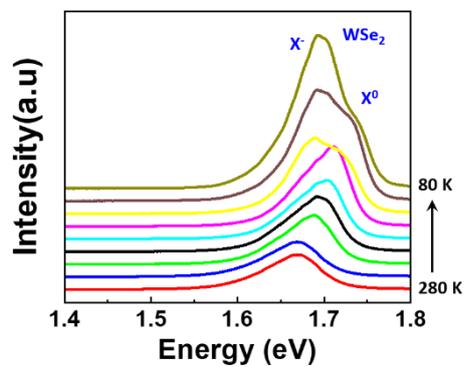

**Figure S10. Temperature-dependent PL spectra (280 K to 80 K) of the heterostructures with 1 min-MoSe₂ QDs.** These spectra were obtained by 633-nm-laser with an excitation power of 20 μW and a 50x lens with 0.35 NA.

**Supplementary Figure S11**

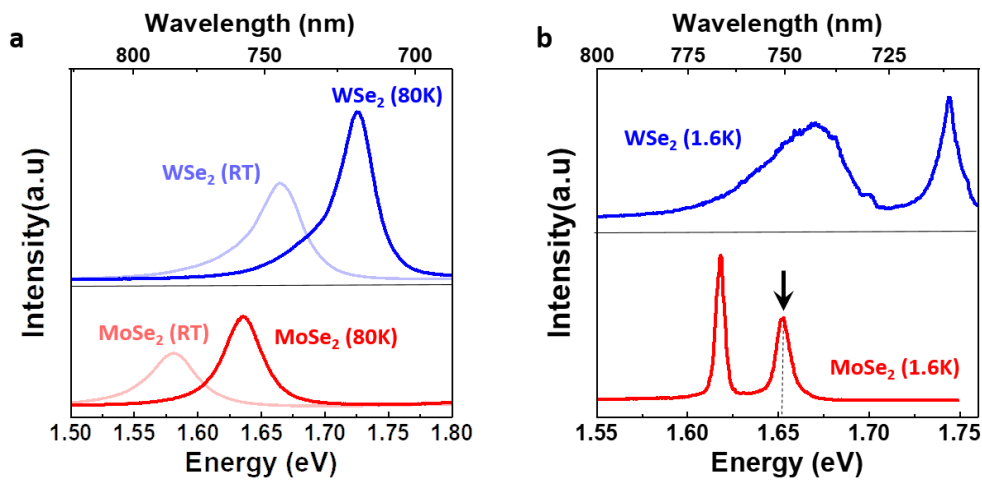

**Figure S11. Temperature-dependent PL spectra of reference MoSe$_2$ and WSe$_2$ monolayers.** (**a**) PL spectra of the WSe$_2$ (top, blue) and MoSe$_2$ monolayers (bottom, red) measured at 80 K (solid) and room temperature (blurry dot) using 633 nm-laser excitation in a Linkam stage with a liquid nitrogen supply. (**b**) Cryogenic PL spectra of WSe$_2$ (top, blue) and MoSe$_2$ monolayers (bottom, red) measured at 1.6 K using 640 nm-laser excitation in a cryostat.

**Supplementary Figure S12**

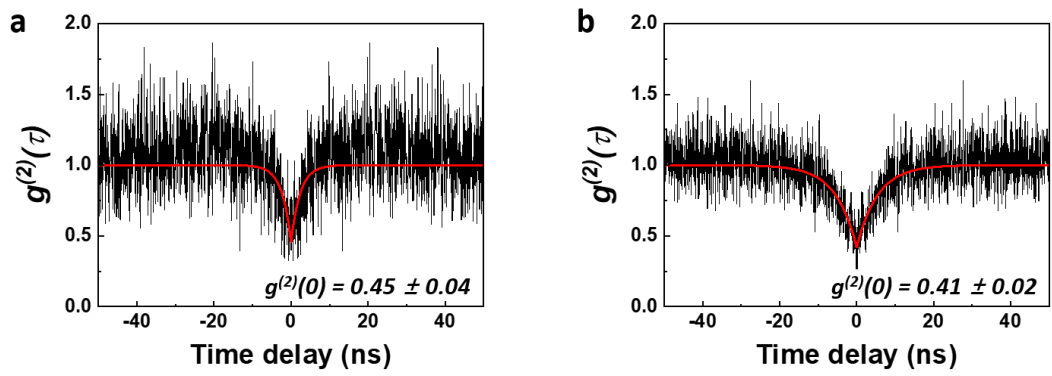

**Figure S12. Additional second-order correlation measurement on other points**